\documentclass[twocolumn,times]{aastex62}
\shorttitle{Rotation History of PSR J1939+2134 (B1937+21)}
\shortauthors{Vivekanand}
\begin{document}
\title{The 31-year Rotation History of the Millisecond Pulsar J1939+2134 (B1937+21)}
\email{viv.maddali@gmail.com}
\author[0000-0002-5498-8988]{M. Vivekanand}
\affil{No. 24, NTI Layout 1\textsuperscript{st} Stage, 3\textsuperscript{rd} Main,
1\textsuperscript{st} Cross, Nagasettyhalli, Bangalore 560094, India.}
\begin{abstract}
The timing properties of the millisecond pulsar PSR J1939+2134 -- very high 
rotation frequency, very low time derivative of rotation frequency, no timing 
glitches and relatively low timing noise -- are responsible for its 
exceptional timing stability over decades. It has been timed by various groups 
since its discovery, at diverse radio frequencies, using different hardware 
and analysis methods. Most of this timing data is now available in the public 
domain in two segments, which have not been combined so far. This work analyzes 
the combined data by deriving uniform methods of data selection, derivation of 
Dispersion Measure (DM), accounting for correlation due to ``red" noise, etc. 
The timing noise of this pulsar is very close to a sinusoid, with a period of 
approximately $31$ years. The main results of this work are (1) The clock of PSR 
J1939+2134 is stable at the level of almost one part in $10^{15}$ over 
about $31$ years, (2) the power law index of the spectrum of electron density 
fluctuations in the direction of PSR J1939+2134 is $3.86 \pm 0.04$, (3) a Moon 
sized planetary companion, in an orbit of semi major axis about $11$ 
astronomical units and eccentricity $\approx 0.2$, can explain the timing noise 
of PSR J1939+2134, (4) Precession under electromagnetic torque with very small 
values of oblateness and wobble angle can also be the explanation, but with 
reduced confidence, and (5) there is excess timing noise of about $8$ $\mu$s 
amplitude during the epochs of steepest DM gradient, of unknown cause.
\end{abstract}

\section{Introduction} \label{sec:intro}

Long term timing of the millisecond pulsar J1939+2134 (henceforth J1939) 
was started by \citet{KTR1994} (henceforth KTR). They describe the method of 
measuring pulse arrival times, estimation of the Dispersion Measure (DM), and 
estimating the timing model used to derive timing residuals, which are the 
final quantity of interest. See \citep{Manchester1977, Backer1986, Lyne2006} 
for pedagogical reviews of pulsar timing.

\subsection{Summary of Pulsar Timing}

The periodic pulses from a pulsar have a polarization that varies through the 
pulse. So they have to be observed using a dual polarization receiver, called
the front-end. This signal travels to earth through the ionized 
interstellar medium (ISM), which causes a frequency dependent delay, that 
depends upon the DM. So a spectrometer is required to divide the total radio 
frequency band into smaller sub-bands, such that the pulse smearing within 
each sub-band is tolerable. Such an instrument is known as the back-end. 
When possible, the total time delay across the total band is used to 
estimate the DM, which is then used to align the total intensity profiles of 
the sub-bands, with respect to that of a reference sub-band. Folding the 
aligned data at the period of the pulsar yields the so called integrated 
profile. The several integrated profiles obtained during a day's observation 
are compared with a template integrated profile, that is specific to each 
pulsar, to derive the average pulse arrival time at the site of the 
observatory for each day; this is known as the site arrival time (SAT).

Often the traditionally used bandwidths do not provide sufficient radio 
frequency separation to estimate the DM accurately. This is particularly 
true for J1939, whose very low period of about $1.56$ milliseconds (ms) requires that 
the SAT be measured 
with accuracies better than $\approx 1$ microsecond ($\mu$s). Therefore 
one needs to observe the pulsar at another well separated radio frequency, 
ideally simultaneously, but often in practice contemporaneously. The 
popular radio frequencies of front-end for pulsar timing are $800$ MHz, $1400$ 
MHz and $2300$ MHz, that fall in 
the microwave frequency bands UHF, L-band and S-band, respectively.

Next the SAT have to be corrected for the delay in the ISM at the frequency 
of the reference sub-band. They also have to be be corrected for solar 
system effects such as the ``Roemer", ``Einstein" and ``Shapiro" delays (see 
KTR and references therein). If the pulsar is in a binary system then
additional corrections are required. This results in the SAT being transformed 
into arrival times at the co-moving pulsar frame. If one ignores constant time 
offsets, this can be considered to be the pulse arrival time at the barycenter 
of the solar system (BAT).

\begin{figure}[b]
\centering
\hspace*{-1.0cm}
\includegraphics[width=1.2\hsize]{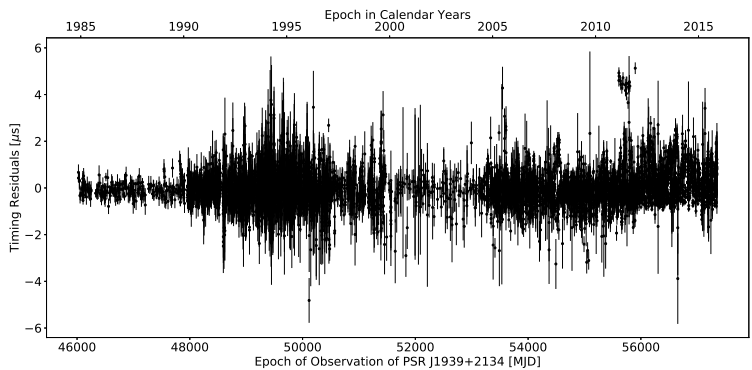}
	\caption{$31$ years of timing residuals of J1939 along with error bars, 
	 after removal of timing noise, which is estimated by TEMPO2 using $100$ 
	 noise harmonics, as explained later in the text. 
        }
\label{fig1}
\end{figure}

Finally, the BAT are modeled using the following parameters: (1) the 
position of the pulsar in the sky (right ascension $\alpha$ and declination
$\delta$), (2) its proper motion ($\mu_{\alpha}$ and $\mu_{\delta}$), (3)
its parallax ($\pi$) and (4) fine correction to the DM, if the data permits 
their modeling, (5) the pulsar's rotation frequency and its derivative with 
respect to time ($\nu$ and $d \nu / d t = \dot \nu$; for J1939 the second 
frequency derivative $\ddot \nu$ is not used as explained later). 
The time difference between the observed and modeled BAT are known as 
timing residuals. In J1939 these residuals represent what is known as timing 
noise (also known as ``red" noise, implying low frequency variation of 
timing residuals). The main effort of this work is to obtain the timing 
noise of J1939. After removal of timing noise, the timing residuals should 
ideally reflect random and uncorrelated noise (mainly instrumental), also 
known as ``white" noise. This is shown in Figure~\ref{fig1}, where the rms 
of the residuals is $\approx 0.5$ $\mu$s. Even if the operative value is $3$ 
times larger, a variation of $\approx 1.5$ $\mu$s over $31$ years implies 
that J1939's clock is stable at the level of almost one part in $10^{15}$. 
This is consistent with the value of one part in $10^{14}$ obtained by KTR 
over about $8$ years of observation.

This brief summary of the technique of pulsar timing ignores several details.

First, the total data consists of about $22$  years of data obtained by European 
and Australian radio observatories, and about $19$ years of data obtained 
by North American radio observatories. The former are known as European Pulsar 
Timing Array (EPTA) data and the Parkes Pulsar Timing Array (PPTA) data. Due 
to instrumentation and methodology differences, the earlier $\approx 8$ years 
of the data from the North American radio observatories are combined with the 
EPTA and PPTA data to form what is known as the International Pulsar Timing Array 
(IPTA\footnote{http://ipta4gw.org//data-release/}) data. The remaining about 
$11$ years of the data are known as the NANOGrav\footnote{https://data.nanograv.org/} 
data. Further, the IPTA \citep{Verbiest2016} and NANOGrav 
\citep{Arzoumanian2018} data of J1939 are obtained by four European and one 
Australian, and two North American, radio telescopes, respectively. Each of 
these observed J1939 for different durations over the last $\approx 31$ years, 
with different front-end/back-end combinations known as sub-systems, that 
changed over time as better sub-systems were installed over time. Now, data 
of any two sub-systems will have a 
relative instrumental delay, usually in the range of $\mu$s to ms, which has 
to be estimated and corrected before the two data can be combined.
There are totally $36$ sub-systems in the J1939 data, so $35$ instrumental
delays have to be estimated to align the whole data set. This is not easy as 
data of different sub-systems often either do not overlap in epoch, or
overlap minimally.

Next, reliable estimation of DM of J1939 requires nearly simultaneous 
observations at two or more radio frequency bands, since the DM
changes with epoch; this is often not achieved. The best data in this regard 
was obtained by KTR in which the dual frequency observations were typically 
separated by about $1$ hour. Often this can be as large as days or even
weeks in the rest of the data. Therefore for some duration, the DM has to be 
modeled (like the other $7$ pulsar parameters) as a function of epoch, 
instead of being directly estimated (which is done using equation $4$ of KTR).

Finally, while modeling the BAT to estimate the various pulsar parameters, 
the presence of ``red" timing noise introduces correlations between adjacent
timing residuals; the correlation length depends upon the ``redness" of the
timing noise. This has to be accounted for in the linear weighted least 
squares parameter estimation algorithm \citep{Coles2011, Caballero2016}.

The IPTA \citep{Verbiest2016} and NANOGrav \citep{Arzoumanian2015, Arzoumanian2018} 
groups analyzed the J1939 data separately, since the two data contain a fundamental 
difference, using different methods of DM estimation, correction for ``red" noise 
correlation, etc.

\subsection{Incompatibility of IPTA and NANOGrav Data}

\begin{figure}[t]
\centering
\hspace*{-1.0cm}
\includegraphics[width=1.2\hsize]{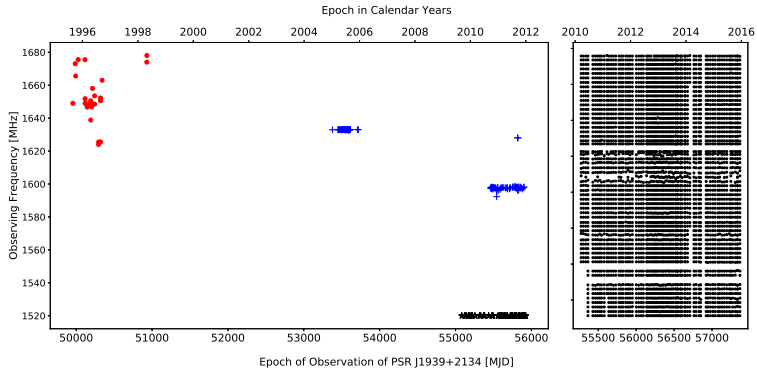}
\caption{Illustration of incompatibility between IPTA and NANOGrav data, using
	 the $1600$ MHz L-band data. The left panel plots the radio frequency 
	 of observation of the IPTA data from Parkes (dots), Jodrell Bank (+) 
	 and Nancay (*) Observatories. The right panel displays the NANOGrav 
	 data from the Green Bank Observatory  that corresponds to the same 
	 frequency range as that in the left panel, although the actual 
	 frequency range of this data is from $1100$ MHz to $1900$ MHz. 
        }
\label{fig2}
\end{figure}

The left panel of Figure~\ref{fig2} shows the exact observational radio frequency 
used to obtain each SAT, as a function of epoch, for three observatories 
of IPTA, viz., the Parkes, Jodrell Bank and Nancay observatories; the observations 
were done at around $1600$ MHz in the L-band. The Parkes data lies at 
frequency $\approx 1650$ MHz, below MJD $51000$; the Jodrell data is 
clustered almost exactly at frequency $1520$ MHz, at MJD between $55000$ and 
$56000$; the rest of the data in the left panel is from Nancay. The Jodrell and 
Nancay data are obtained within a relatively narrow band of frequencies around 
the central frequency. The central frequency of the Nancay data changes slightly 
between the earlier and later epochs, ignoring two points which appear to be 
outliers.  The Parkes data is more spread out in frequency, but still within a 
band of $\pm 30$MHz around the central frequency. Moreover, the IPTA data 
consists of just one SAT at each epoch of observation.

In contrast, the NANOGrav data from Green Bank Observatory in the right panel is 
spread over a band 
of about $160$ MHz. Moreover, the NANOGrav data consists of several tens of SAT
(sometimes as large as $50$) at each epoch of observation. This is because of 
the extremely wide (radio) band sub-systems used at the Green Bank Observatory. 
Since the IPTA and NANOGrav data overlap in epoch, they have been plotted in 
separate panels in Figure~\ref{fig2} for clarity. The situation depicted 
in this figure holds true for the data in other radio frequency bands as well.

Now, it is well known that the integrated profile of pulsars is radio frequency 
dependent, and usually becomes narrow at higher frequencies \citep{Manchester1977}. 
This would cause an additional frequency dependent delay in the SAT, that is 
not DM related. This aspect can be ignored when the observing bandwidth is 
narrow (IPTA data), but can not be for wide-band sub-systems (NANOGrav data). The 
NANOGrav group has dealt with this issue by introducing what are known as ``FD" 
parameters into their analysis; see section $4.2$ of \citep{Arzoumanian2015}. 
Now, using the FD parameters for the IPTA data would distort the timing residuals.
Therefore there is a fundamental issue involved in combining the IPTA and NANOGrav 
data for analysis.

\cite{Verbiest2016} combine such data for other pulsars, and discuss the problems
involved, but not in the manner described here; see particularly their sections 
$1.4$, $2.2$ and $3.1$.

\subsection{The Current Work}

This work combines the two data sets by selecting only a few SAT per epoch of the NANOGrav
data, such that they lie within a narrow radio bandwidth, and also such that their 
spread in the time domain is less than $1$ $\mu$s.
This eliminates the need for the FD parameters. Tests show that this number can 
be as low as one SAT per epoch, mainly on account of the excellent quality of the 
NANOGrav data.

Section $2$ describes the observations and the procedure of analysis, including 
data selection and estimation of the DM. Section $3$ presents the results of 
analyzing the data in four independent ways,  while section $4$ discusses the
results. 

\section{Observations and Analysis}

\begin{figure*}[t]
\centering
\includegraphics[width=\hsize]{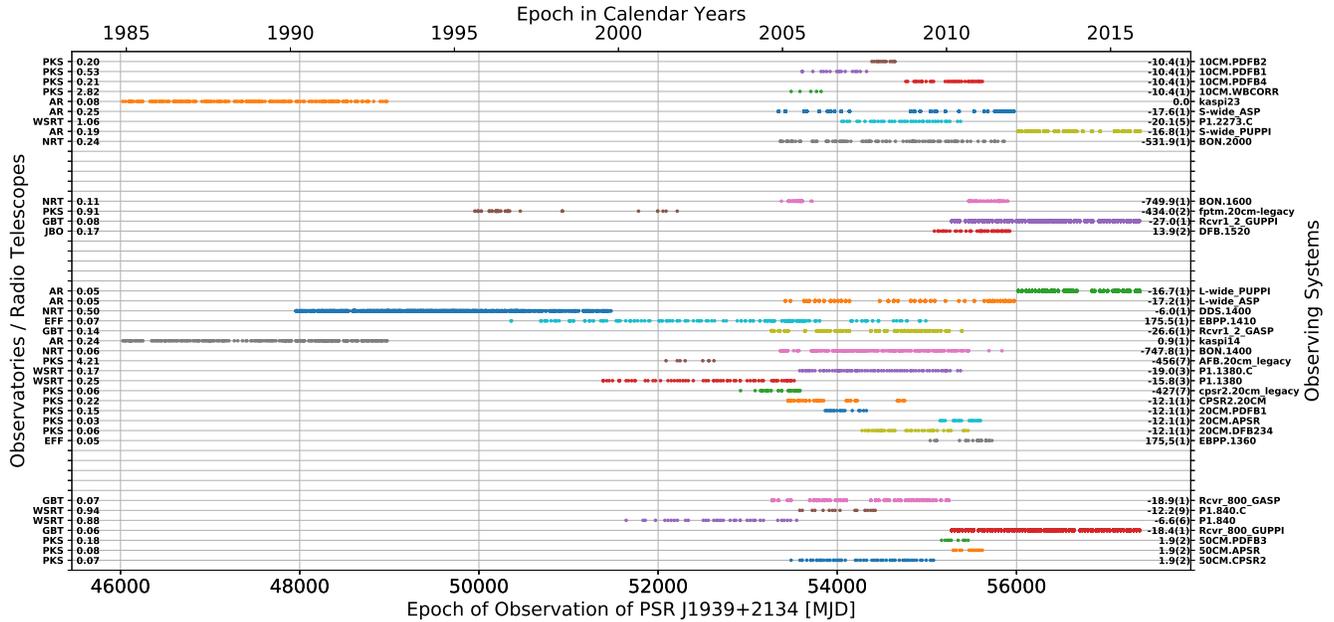}
\caption{Summary of IPTA and NANOGrav observations of J1939 that are available in the
	 public domain. The first column of labels of the left ordinate are: (1)
	 AR = Arecibo Telescope, (2) EFF = Effelsberg Telescope, (3) GBT = Green 
	 Bank Telescope, (4) JBO = Jodrell Bank Observatory, (5) NRT = Nancay 
	 Radio Telescope, (6) PKS = Parkes Telescope, and (7) WSRT = Westerbork
	 Synthesis Radio Telescope. The second column of labels are the average
	 error on the SAT in $\mu$s.  The first column of labels of the right 
	 ordinate are the instrumental offsets (in $\mu$s) for each sub-system, 
	 with respect to that of ``kaspi23";  the error in the last digit is given
	 in parenthesis. The second column of labels 
	 of the right ordinate represent sub-systems. Some important references for 
	 this table are: \cite{Arzoumanian2015, Arzoumanian2018, Backer1982, 
	 Cordes1990, Hotan2006, KTR1994, Manchester2013, Ramachandran2006, 
	 Shannon2010, Verbiest2009, Verbiest2016}, and references therein.
        }
\label{fig3}
\end{figure*}

Figure~\ref{fig3} summarizes the available data of J1939. Each point represents an
observation of either single or multiple SAT at that epoch. Each horizontal 
track represents data of one sub-system, indicating both the total duration of 
observation, as well as the cadence of observation in that duration. The bottom
$7$ tracks belong to the UHF band, the middle $20$ belong to the L-band, and the 
top $9$ belong to the S-band; space is provided between bands for clarity. For
further clarity, the $16$ tracks at frequency $\approx 1400$ MHz in the L-band 
are separated by space from the $4$ tracks at frequency $\approx 1600$ MHz in 
the same band. The left ordinate is labeled by the observatory or radio 
telescope concerned, and the average statistical error of the SAT of that track 
(in $\mu$s). The right ordinate is labeled by the symbol used for the observing 
sub-system by the original observers,  and their relative  instrumental 
offsets in $\mu$s. 

The four L-band tracks at $1600$ MHz are expanded in frequency in 
Figure~\ref{fig2}, except for that of the Parkes Telescope (PKS) 
sub-system ``fptm.20cm-legacy", which consists of both $1400$ and $1600$ MHz
data; so only the latter part has been used in Figure~\ref{fig2}.

The earliest observations of J1939 are those of KTR using the Arecibo Telescope 
(AR) (sub-systems kaspi14 and kaspi23 in the L and S bands; see the caption of 
Figure~\ref{fig3} for observatory abbreviations). Then Nancay Radio Telescope 
(NRT)  started observing in just the L-band using an older sub-system (DDS.1400). 
This provided relatively lower quality data, but it was their first sub-system, and 
provided crucial overlap with the data of KTR; their later sub-systems provided data as 
good as any other sub-system. Then the Effelsberg Telescope (EFF) started 
observations with 
overlap with the data of NRT. Their duration of observation was one of the 
longest, although there were gaps in the observations; and their data 
quality is one of the best; unfortunately theirs was also a single frequency 
observation (L-band). Since after the year $2011$ it appears that 
data of J1939 is being provided by only the North American telescopes.

Figure~\ref{fig3} shows the inhomogeneity of the data of J1939, in terms of 
duration of observation, the cadence of observation within any duration, and 
the quality of the data. It also shows that for about $7$ years after MJD 
$49000$, there were no multi-frequency observations available
in the public domain that could be used to estimate the DM.

\subsection{Data Selection}

\begin{figure}[t]
\centering
\hspace*{-1.0cm}
\includegraphics[width=1.2\hsize]{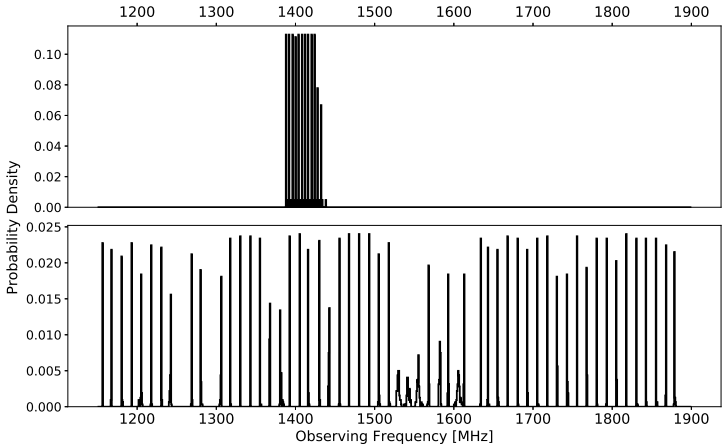}
\caption{Probability distribution of frequency of observation for the 
	Rcvr1\_2\_GASP (top panel) and Rcvr1\_2\_GUPPI (bottom panel)
	sub-systems, respectively.
        }
\label{fig7}
\end{figure}

The rationale of data selection can be understood using Figure~\ref{fig7}.
The narrow band sub-system Rcvr1\_2\_GASP of Green Bank Telescope (GBT), 
of bandwidth of about $50$ MHz, was replaced by the broad band sub-system 
Rcvr1\_2\_GUPPI, of bandwidth of about $740$ MHz, towards the end of the 
year $2010$. Because of frequency evolution, the pulse of J1939 arrives at 
different times at different frequencies within the band of observation, 
even after removing the effects of DM. Ideally one should have a unique 
template integrated profile at each frequency for obtaining the SAT at that 
frequency, but this is too humongous a task.  Therefore the integrated 
profile at a reference frequency within the band of observation is used as 
the template for the entire band. Now it turns out that the frequency 
evolution of the pulse profile is negligible for the narrow band sub-system, 
while it is significant for the broad band sub-system.

If the broad band sub-system was not installed, and the narrow band 
sub-system had continued observing J1939, then the problem of frequency 
evolution would not have arisen. In this work such a hypothetical
scenario is created, by using from the broad band sub-system, only that data that 
corresponds to the range of frequencies of the narrow band sub-system. 
There are several ways of doing this, and this work adopts one of those.

This implies that one would be excising most of the broad band data. This 
is justified because an observation using a wider band improves the signal 
to noise ratio of the integrated profile only if the pulses arrive at the 
same time all over the band, which is not the case here. This work 
demonstrates that the large collecting areas of the GBT and AR telescopes 
ensure that there is sufficient signal to noise ratio of the integrated 
profile within the retained narrow band, to obtain a statistically 
significant SAT. This scheme will probably not work with wide band data 
from smaller radio telescopes.

The data selected for analysis in this work consists of the entire IPTA data, 
and part of the NANOGrav data, which was selected as follows. The NANOGrav 
data itself consists of two relatively narrow band sub-systems labeled ASP 
and GASP, and two very broad band sub-systems labeled PUPPI and GUPPI; ASP 
and PUPPI are the back-ends used at the AR telescope, while GASP and GUPPI are 
identical back-ends used at the GBT.
The mean radio frequency of the data of ASP and GASP systems in the UHF, L-band 
and S-band are $844$, $1410$ and $2352$ MHz, respectively. At each epoch of 
observation, say $n$ SAT were selected from the multiple SAT available, that 
were closest in frequency to any of the above three values. For small values of 
$n$ the selected SAT would have very narrow spread in frequency, and would have 
very small systematic spread in time due to frequency evolution of the pulse 
of J1939; for large $n$ the situation would be the opposite. In either case the
$n$ SAT would have a mean frequency very close to one of the above three values. 
Several values of $n$ were tried, and values between $1$ and $10$ were found to be 
useful. Even $n = 1$ served the essential purpose, presumably because of the 
excellent quality of NANOGrav data. However $n = 7$ to $10$ were found to be better 
for DM estimation; therefore $n = 10$ was finally chosen. Larger $n$ caused the
systematic spread in time of the SAT to be larger than $1$ $\mu$s. This selection 
reduced the total data from $18122$ to $6994$ SAT.

In summary, frequency evolution of the pulse essentially converts a broad band 
observation of J1939 into several narrow band observations that are each observing, 
in effect, a slightly different pulsar, of which one has been chosen in this work.

\subsection{Estimation of DM}

The DM of J1939 has to be estimated as a function of epoch to proceed further.
The best DM estimate has been done by KTR from MJD about $46000$ to MJD 
$49000$.  From then on until MJD $51650$ only single frequency 
observations are available in the public domain. However, \cite{Ramachandran2006} 
used additional data, obtained at several frequencies using the NRAO $85$ and 
$140$ foot telescopes at the Green Bank Observatory, to extend the DM measurements to slightly 
beyond calendar year $2004$. While this additional data of J1939 are not 
available in the public domain, the DM results are available in Figure~$6$ of 
\cite{Ramachandran2006}, and also in Figure~$6.6$ of \cite{Demorest2007}, from 
which they were digitized. Next one has to estimate the DM of J1939 for the
rest of the data, and align it with the above digitized curve.

The ideal method of estimating the DM at any epoch is to measure the SAT at 
two well separated frequencies simultaneously, and then to apply equation 
$4$ of KTR. Such SAT would measure the arrival of exactly the same pulse, but at 
different frequencies.  However this is rarely possible, since changing the 
front-end of a radio telescope takes time. Therefore the method of KTR is the next 
best, where the dual frequency observations are separated by about $1$ 
hour. Since the maximum rate of change of DM of J1939 is about 
$10^{-5}$ pc cm$^{-3}$ per day (as will become evident later), and 
since the error on the estimated DM is typically larger than $10^{-4}$ 
pc cm$^{-3}$, one can assume that a gap of even $10$ days between multi 
frequency observations is tolerable for DM estimation of J1939. The IPTA 
group appears to have used some variation of the KTR method, and have 
additionally modeled the first and second derivative of DM with respect to 
epoch. The data gap between NANOGrav observations is much longer, typically 
$10$ to $14$ days. Therefore they have modeled the DM along with other 
pulsar parameters (as a function of epoch) using the so called ``DMX" parameters.

The approach of this work is to apply the KTR method, allowing for a
maximum gap of about $1$ day between multi frequency observations
(see \cite{Lam2015} for justification). This
reduces the number of epochs at which the DM can be estimated, but linear
interpolation for the intermediate epochs gives satisfactory results.
Equation $4$ of KTR can be re-written as consisting of a term that varies 
linearly with DM and inversely with the square of the observing frequency, 
plus a constant term that represents instrumental and other 
constant delays. Thus for dual frequency data one has to model for two 
parameters -- the DM and one constant relative delay between the two 
frequencies; for three frequency data one has to model for the DM and two 
constant relative delays. In this manner the DM as a function of epoch 
was derived separately for each radio telescope, along with the relative 
instrumental offsets for the corresponding frequencies. These curves were 
aligned in the DM space, and the result was then aligned with the digitized 
DM curve from \cite{Demorest2007}.

The TEMPO2 software \citep{Hobbs2006, Edwards2006} has been used for most 
of the analysis of this work.

The DM in this work was estimated by first pruning the data of each 
telescope, such that only those SAT were retained that had at least one 
other SAT at another frequency band, that was separated in epoch by less than one 
day. The reference epoch for the fit (PEPOCH in TEMPO2) was taken to be 
mid way between the total duration of the pruned data. Now, TEMPO2 provides 
for inserting time offsets between data sets using the ``JUMP" parameter. 
So JUMP values were inserted for each sub-system within the UHF, S-band and 
1600 MHz data of the L-band, with respect to the 1400 MHz data of the 
L-band data. A constant value of $=71.0270$ pc cm$^{-3}$ was used 
for the DM parameter of TEMPO2. Correlation due to ``red" noise was taken into 
account, but was required only for the WSRT telescope, whose data duration
was quite long; for the rest of the telescopes ``whitening" of the data
was achieved by using the $\ddot \nu$ parameter in TEMPO2. Note that
this $\ddot \nu$ represents the local curvature of the timing data for
each telescope -- it is not related to the intrinsic $\ddot \nu$ of 
J1939, which is too small to be estimated in our data. Then the residuals of 
the fit were extracted using the ``general2" plugin of TEMPO2. These were 
then analyzed outside TEMPO2 to estimate the residual DM as a function of 
epoch, using Equation 4 of KTR. The sum of the residual DM and $71.0270$ 
results in the final DM as a function of epoch, for each telescope.
Note that TEMPO2 has the ability to estimate the instrumental offset of
data that do not overlap in epoch, as long as the lack of overlap is not
of very long duration, and as long as the data is sufficiently ``whitened".

\subsection{Analysis of SAT}

A uniform set of TEMPO2 parameters was used
for the selected data; the IPTA and NANOGrav groups used 
different values for these parameters (see Appendix A).

The DM for each SAT was estimated using linear interpolation on the final 
DM curve obtained in the previous section; this was tagged to each SAT 
using the TEMPO2 flag ``-dmo", after removing the original DM tag inserted 
by the IPTA group. No further DM modeling was done in TEMPO2. 

Next, the L-band $1400$ MHz data of each telescope was aligned, by estimating 
the instrumental delays between them, using the method described in the 
previous section. The instrumental delays of the UHF and S-band data of each 
telescope, relative to their $1400$ MHz L-band data, that were estimated in 
the previous section, were then used to align the rest of the data; only 
minor changes were required in these delay values for final data alignment. 
Similarly the L-band $1600$ MHz data were also aligned. The IPTA and 
NANOGrav groups started with initial JUMP values, and then varied them 
as any other parameter to be fit.
This was tried in this work, but better results were obtained by keeping 
them fixed; so the JUMP values in this work were first estimated as well as
possible, and then were held fixed in TEMPO2 (see Appendix A of 
\cite{Arzoumanian2015}). Some observatories have fixed instrumental delays
for some SAT, that were included without modification in this work.

Next TEMPO2 is used to derive the values of the parameters of the timing model 
-- $\alpha$, $\delta$, $\mu_{\alpha}$, $\mu_{\delta}$, $\pi$, $\nu$ and 
$\dot \nu$. For J1939 the $\ddot \nu$ is expected to be so small (based on
the average braking index of a pulsar) that it can not be reliably estimated
in this analysis. If the timing residuals after accounting for this model 
represent ``white" noise, then the above parameters would have been estimated 
in a reliable manner. However, it is known for J1939 that the residuals are 
slowly varying and sinusoidal (see Figure $5$ of \cite{Verbiest2016}). 
Therefore, the correlation of this ``red" noise has to be taken into account 
for parameter estimation in TEMPO2. This has been done differently by the 
IPTA and NANOGrav groups.

The IPTA group models this correlation as a specific function, and estimates
the three parameters of this function from the data, and inputs these 
three parameters to TEMPO2 (see \citet{Lentati2015} for some details). The 
NANOGrav group models this correlation as arising from a power law spectrum, and 
inputs the amplitude and slope of this spectrum to TEMPO2, which then derives 
$100$ harmonics whose spectrum is (or at least should be) the above power law, 
and which estimate the timing noise. See section $5.1$ of 
\citep{Arzoumanian2015} for details. In this work both methods are used.

The TEMPONEST software \citep{Lentati2014} was installed, but could not be used
on account of prohibitively long run time on my personal computer for $6994$
SAT with ``red" noise covariance included. However, the 
MULTINEST\footnote{https://github.com/JohannesBuchner/MultiNest.git} software
\citep{Feroz2008, Feroz2009, Buchner2014} for Markov Chain Monte Carlo (MCMC) 
estimation of parameters has been used \citep{Hogg2010, Hogg2012, Hogg2018}.

Further details of analysis are given in the following section.

\section{Results}

\subsection{Dispersion Measure of J1939}

\begin{figure}[t]
\centering
\hspace*{-1.0cm}
\includegraphics[width=1.2\hsize]{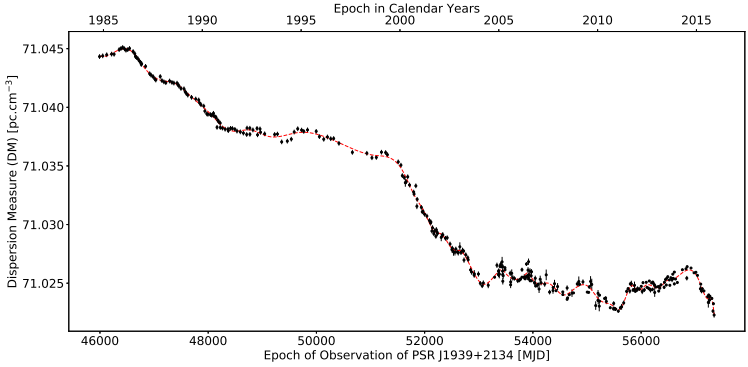}
	\caption{DM of J1939 as a function of epoch for the combined IPTA and 
	NANOGrav data. The earlier DMs have a fixed error of $3 \times 10^{-4}$ 
	pc cm$^{-3}$, which is explained in the text; the later DMs have 
	estimated errors. The dashed line is a spline curve that best fits the 
	data. The maximum gradient is $10^{-5}$ pc cm$^{-3}$ per day.
        }
\label{fig4}
\end{figure}

Figure~\ref{fig4} shows the DM derived for J1939 from the combined IPTA and NANOGrav data.
It is almost identical to that of KTR for the first $8$ years, and correlates very 
well with that of NANOGrav \citep{ Arzoumanian2018} for the last 
$11$ years. The dashed curve is a smooth representation of the DM data,
obtained using the ``splinefit" tool of the open source software 
``octave". The rms of the residuals between the DM data and the spline curve is $3 
\times 10^{-4}$ pc cm$^{-3}$, which is similar to the uncertainty in aligning the 
DM data of different telescopes. 

The spline curve was used to obtain the phase structure function $D_\phi(\tau)$ of 
DM variations, that is defined in equation $11$ of KTR, using equations A2 and A3 
of \cite{You2007}. The actual DM data can not be used for this purpose because the 
earlier half of DM data have no error bars available; these are required to subtract a 
bias in the function $D_\phi(\tau)$. The power law index of the spectrum of 
electron density fluctuations $\beta$ is $3.86 \pm 0.04$ using $31$ years 
of data.  This is consistent with the value of $3.874 \pm 0.011$ obtained by KTR, 
who used the first $8$ years of data. \cite{Ramachandran2006} obtained a 
slightly smaller value of $3.66 \pm 0.04$ using the first $20$ years of 
data. It is therefore concluded that phase structure function $D_\phi(\tau)$ of 
DM variations of J1939 is not evolving over the decades.


\subsection{Timing Noise of J1939}

Both IPTA and NANOGrav groups use what are known as the T2EFAC and T2EQUAD parameters 
(henceforth T2 parameters), one pair for each sub-system. The former is used to 
scale the measured uncertainties on the SAT, while the latter is added to them 
in quadrature. These are used to ensure that the final reduced $\chi^2$ obtained 
by TEMPO2 is close to the expected value of $1$; see section $3.1.2$ of 
\cite{Verbiest2016}. In addition the NANOGrav group uses the ECORR (or jitter) 
parameter, also one for each sub-system, that acts like the T2EQUAD for data 
spread in frequency (see section $4.2$ of \cite{Arzoumanian2014}, and section 
$3$B of \cite{vanHaasteren2014}). In this work, firstly the ECORR parameter is 
not used, on account of the data selection discussed above. Next, the analysis 
was done using both the original values of the T2 parameters (derived by the 
IPTA and NANOGrav groups), as well as those that were re-estimated here using the 
``fixData" plugin of TEMPO2. The latter were consistent with the former, although 
for some sub-systems the values differed significantly.

\begin{figure}[b]
\centering
\hspace*{-1.0cm}
\includegraphics[width=1.2\hsize]{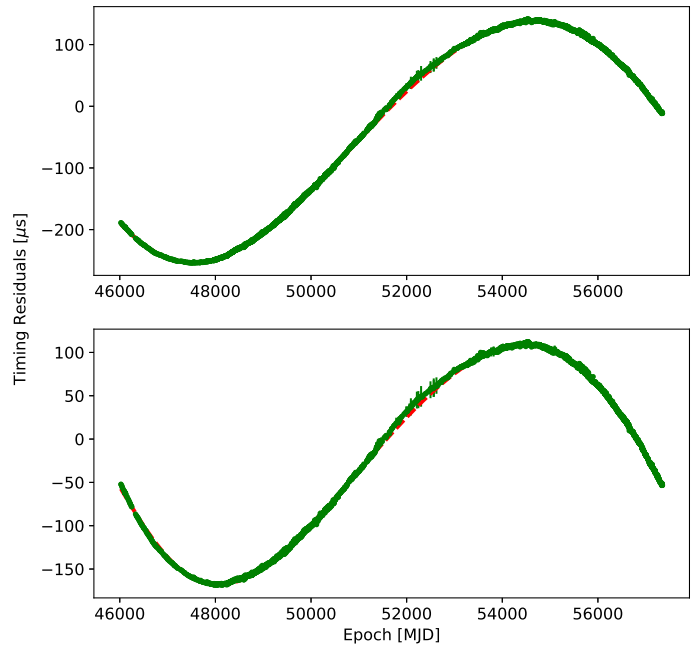}
\caption{The data in the two panels were obtained using the IPTA (top panel)
	and the NANOGrav (bottom panel) methods of correction for ``red" 
	noise correlation; in both cases re-estimated T2 parameters were 
	used. The dashed line in each panel represents the best fit 
	planetary companion model. The IPTA group models the ``red" 
	noise correlation as a specific function with three parameters; these are
	given in the last three rows of Table~\ref{tabl1}. The NANOGrav 
	group models the power spectrum of this correlation as a power law,
	with amplitude and slope of $0.15726$ and $-2.7589$, respectively. TEMPO2 
	uses these to derive $100$ harmonics which represent the timing noise. 
	}
\label{fig5}
\end{figure}

In the following sections the timing noise is modeled as being on account of 
(1) a planetary companion to J1939, and (2) precession of J1939. In principle 
the former model can be explored using the binary parameters of TEMPO2. 
However, this attempt failed due to either (a) converging to negative values 
of eccentricity of the elliptical orbit, or (b) resulting in very large
errors on the binary parameters, both presumably because the data has only 
one cycle of the orbit. Therefore the results of the following sections are 
derived by using independent software.

In principle, one can not rule out a third model, viz., that the timing 
noise could be due random variations in the frequency $\nu$, the variations 
having a steep power law spectrum. It is argued in later sections that this 
may not be the case.

\subsubsection{Planetary Companion Model}

Table~\ref{tabl1} summarizes the results of this section while Figure~\ref{fig5}
illustrates two of them (columns $3$ and $5$) as plots. In the top panel of
Figure~\ref{fig5}, the noise model was determined from the data using the 
``autoSpectralFit" plugin, and input to TEMPO2 using the ``-dcf" switch; the 
parameters of the model (slope $\alpha$, cutoff frequency $f_c$, and 
amplitude $a$) are given in the last three rows of columns $3$ of 
Table~\ref{tabl1}. In the bottom panel, the NANOGrav noise model was used -- 
RNAMP $= 0.15726$ and RNIDX $= -2.75890$ (see Table $3$ and Figure~$3$ of 
\cite{Arzoumanian2015}).

In Figure~\ref{fig5} there is an excess timing noise between MJD $51200$ 
and $53200$; this coincides with the duration of maximum gradient of
DM in Figure~\ref{fig4}. These data have been ignored during the curve fit, 
and will be discussed later.

\begin{table}[t]
\scriptsize
\setlength\tabcolsep{1.5pt}
\hspace*{-1.0cm}
\caption{Results of fitting the planetary companion model to the timing noise, 
	that was obtained using both the IPTA (columns $2$, $3$) and the 
	NANOGrav (columns $4$, $5$) methods of correction for ``red" 
	noise correlation. In each case, both the original T2 parameters (orig), 
	as well as those re-estimated here (local), were used.  $A$ is the 
	projected semi major axis of the orbit, $e$ is its eccentricity and $P$ 
	is the period of orbit. The last three rows of IPTA 
	contain the derived parameters of the noise model  -- $\alpha$ and $a$ are
	the exponent and amplitude, while $f_c$ is the cut-off frequency. For the 
	NANOGrav noise model, the original RNAMP $= 0.15726$ and RNIDX $= 
	-2.7589$ were used.
	\label{tabl1}}
\begin{tabular}{lcccc}
\hline
\hline
{Method}  & \multicolumn{2}{c}{IPTA} & \multicolumn{2}{c}{NANOGrav} \\
\hline
{T2param}  & {$orig$}  & {$local$} & {$orig$}  &  {$local$}  \\
\hline
{$A$} ($\mu$s) & $129.9 \pm 0.1$ & $197.6 \pm 0.1$ & $140.6 \pm 0.1$ & $141.1 \pm 0.1$ \\
{$e$} & $0.219 \pm 0.001$ & $0.158 \pm 0.001$ & $0.211 \pm 0.001$ &  $0.215 \pm 0.001$ \\
{$P$} (days) & $11105 \pm 2 $ & $13068 \pm 3$ & $11403 \pm 2$ & $11381 \pm 2$ \\
{rms} ($\mu$s) & $1.5$ & $1.0$ & $1.2$ &  $1.2$ \\
{$\chi^2_d$} & $35.0$ & $30.2$ & $24.0$ &  $39.2$ \\
{$\alpha$} & $3.016$ & $5.098$ & & \\
{$f_c$} & $0.0322$ & $0.0322$ & &  \\
{$a$} & $6.4\times10^{-23}$ & $9.8\times10^{-20}$ &  &  \\
\end{tabular}
\end{table}

Table~\ref{tabl1} show the results of four different analysis -- using the
IPTA and the NANOGrav methods of correction for ``red" noise  correlation, and 
in each of these, using original T2 parameters as well as re-estimated ones.  
The timing noise in columns $2$ to $5$ of Table~\ref{tabl1} are fit to a 
planetary companion model. The parameters of the elliptical orbit are: the 
projected semi major axis of the orbit $A$, its eccentricity $e$ and period 
of orbit $P$, epoch $T0$ and longitude $\omega$ of periastron; only the 
important first three parameters are listed in the first three rows of 
Table~\ref{tabl1}. The formula for the BAT in this case is well known (for 
example see Eq.~$9$ of \cite{Malhotra1993}). No approximation has been made 
in fitting the planetary model -- the full Kepler equation has been solved.
The results of Table~\ref{tabl1} have been obtained using the ``curve\_fit"
tool of the Python module ``scipy". Then they have been verified, 
particularly regarding the distribution of errors and their correlations,
using the ``solve" tool of the Python implementation of MULTINEST 
(pymultinest\footnote{https://github.com/JohannesBuchner/PyMultiNest}). 
To speed up this algorithm the critical code was written in 
C, and called as a library in Python using its C interface.

The parameters of the top and bottom panels of Figure~\ref{fig5} are given in 
columns $3$ and $5$ of Table~\ref{tabl1}. $rms$ is the standard deviation 
of the timing noise after subtraction of the planetary companion model, 
while $\chi^2_d$ is the $\chi^2$ per degree of freedom of the fit. In all 
four cases the $\chi^2_d$ is much higher than $1$, indicating that 
the formal uncertainties are underestimated even after implementing the T2 
parameters. However the $rms$ is less than $1.5$ $\mu$s, which is a small 
fraction of the total amplitude $A$ of the fit.

The timing residuals in Figure~\ref{fig1} are the difference between the timing 
noise in column $4$ of Table~\ref{tabl1}, and the curve represented by the 
$100$ noise harmonics. this is not the same as subtracting the planetary
companion model. The $100$ noise harmonics model fits almost every twist and 
turn of the timing noise, including the excess timing noise between MJD $51200$ 
and $53200$.

The NANOGrav analysis yields consistent values of $A \approx 140$ $\mu$s, $e 
\approx 0.21$ and $P \approx 11400$ days. The IPTA analysis using original
T2 parameters is also consistent with the above values (column $2$ of 
Table~\ref{tabl1}); only the results in column $3$ are divergent. In the 
rest of this section the average of the above three values will be adopted,
viz., $A = 137.2$ $\mu$s, $e = 0.215$ and $P = 11296$ days.

Using the above $A$ and $P$ values in Eq~$2$ of \cite{Vivekanand2017}, the
mass of the planetary companion is about $3.5 \times 10^{-8}$ times the 
solar mass, which is approximately the mass of the moon. Using Kepler's third 
law the semi-major axis of the relative orbit is about $11$ AU 
\citep{Starovoit2017}.

\subsubsection{Precession Model}

The model of a freely precessing pulsar can be understood using Figure $3$
of \cite{Link2001}. The angular momentum and dipole moment vectors of J1939
make the angles $\theta$ and $\chi$ with its symmetry axis, respectively. 
 However since pulsars slow down due to electromagnetic torque, J1939 
must be precessing under the influence of a torque. In this scenario the 
timing residuals are given by:

\begin{equation}
f(t) = k + a_1 \sin \left (\omega_p (t - t_0) \right ) - a_2 \sin 2 
	\left (\omega_p (t - t_0) \right ),
\end{equation}
\noindent where for small wobble angle $\theta$
\begin{equation}
a_1 = \frac{\kappa \theta \sin 2 \chi}{1 + \theta^2}; 
a_2 = \frac{\kappa \theta^2 \sin^2 \chi}{4(1 + \theta^2)}.
\end{equation}
\noindent (Eq.~C46 of \cite{Akgun2006}; see also Eq~$13$ of \cite{Link2001}, and 
also \cite{Jones2001}).  $k$ is an arbitrary offset, $a_1$ and $a_2$ are the 
amplitudes of the first and second harmonics of  the precession frequency 
$\omega_p = 2 \pi \nu_p = 2 \pi / P $. $\kappa$ is proportional to the strength 
of the spin down torque of J1939.

\begin{table}[b]
\setlength\tabcolsep{1.5pt}
\hspace*{-1.0cm}
\caption{Results of fitting the precession model to the timing noise; the rest
	is as in Table~\ref{tabl1}, except that the last three rows of that
	table are not repeated here. $a_1$ and $a_2$ are the amplitudes of 
	the first and second harmonics of period $P$ (see Equation $1$).
	\label{tabl2}}
\begin{tabular}{lcccc}
\hline
\hline
{Method}  & \multicolumn{2}{c}{IPTA} & \multicolumn{2}{c}{NANOGrav} \\
\hline
{T2param}  & {$orig$}  & {$local$} & {$orig$}  &  {$local$}  \\
\hline
{$a_1$} ($\mu$s) & $124.4 \pm 0.1$ & $193.2 \pm 0.1$ & $135.0 \pm 0.1$ &  $135.2 \pm 0.1$ \\
{$a_2$} ($\mu$s) & $9.9 \pm 0.1 $ & $10.2 \pm 0.1$ & $11.0 \pm 0.1$ & $11.2 \pm 0.1$ \\
{P} (days) & $11277 \pm 3$ & $13475 \pm 3$ & $11594 \pm 3$ &  $11609 \pm 4$ \\
{rms} ($\mu$s) & $3.4$ & $2.5$ & $2.9$ &  $2.9$ \\
{$\chi^2_d$} & $174.8$ & $104.1$ & $141.2$ &  $225.9$ \\
\end{tabular}
\end{table}

\begin{figure}[t]
\centering
\hspace*{-1.0cm}
\includegraphics[width=1.2\hsize]{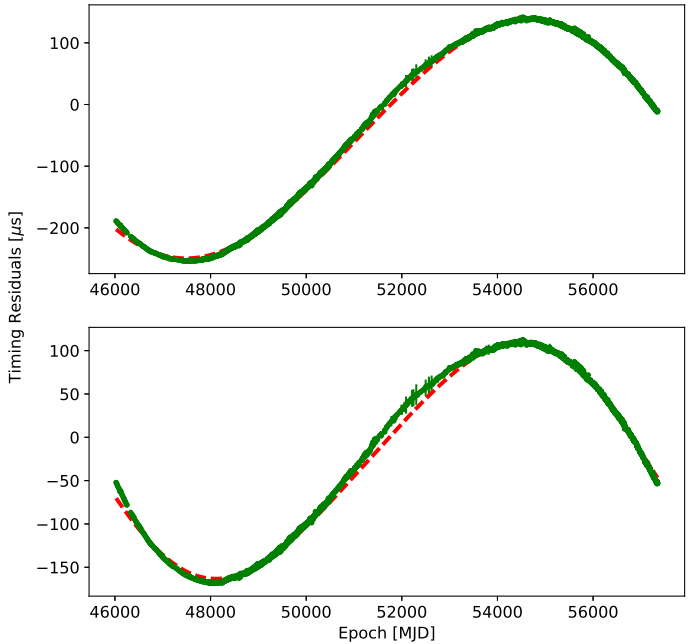}
	\caption{The data is the same as in Figure~\ref{fig5}. The dashed line in 
	each panel represents the best fit precession model.
        }
\label{fig6}
\end{figure}

Table~\ref{tabl2} summarizes the results of this section while Figure~\ref{fig6}
illustrates two of them (columns $3$ and $5$) as plots. Figure~\ref{fig6} is the
same as Figure~\ref{fig5} except that the curve fitted to the timing noise is
given in Eq~$1$. Only the important last three parameters of Eq~$1$ are listed 
in the first three rows of Table~\ref{tabl2}.

The $\chi^2_d$ in Table~\ref{tabl2} are $\approx 5$ times larger than those in
Table~\ref{tabl1}, while the $rms$ are $\approx 2.5$ times larger. Clearly the
data fits the planetary companion model better than the precession model. This 
is also obvious when comparing the solid curves  in Figure~\ref{fig5} and 
Figure~\ref{fig6}. As before, the average of the values in columns $2$, $4$ and 
$5$ of Table~\ref{tabl2} will be used for further analysis, viz., $a_1 = 131.5$ 
$\mu$s, $a_2 = 10.7$ $\mu$s, $P = 11493$ days.

The oblateness of J1939 $\epsilon = (I_3 - I_1)/I_1$, where $I_i$ are the three 
components of the moment of inertia (for a bi-axial rotator $I_3 > I_2 = I_1$). 
Within the approximations used to derive Eq~$1$, the oblateness is $\epsilon 
\approx \omega_p / \omega = \nu_p / \nu = 1 / (11493 \times 86400) / 641.928 
\approx 1.57 \times 10^{-12}$; see also Eq~$67$ of \cite{Jones2001}.

Using Eq~$2$ above $\theta \tan \chi = 8 \times a_2 / a_1 = 0.65$ radians. 
Integration of Eq~$19$ of \cite{Link2001} gives $\theta \tan \chi = 2 \times 
a_2 / a_1 = 0.16$ radians, which is similar to the above result correct to 
within a factor of $4$. However, using Eq~$65$ of \cite{Jones2001}, $\theta 
\approx 3 \times 10^{-6} \tan \chi$ radians, giving an altogether different
functional form. Therefore there is some discrepancy in the values of 
$\theta$ obtained by the three groups (\citet{Akgun2006, Link2001, Jones2001}),
although all three formulae are derived under similar approximations. For this
work $\theta \tan \chi \approx 0.4$ radians will be assumed. Since $\chi$ in J1939 is 
supposed to be very close to $90^{\arcdeg}$ (it has an inter pulse), $\theta$ 
is expected to be a very small value.

\subsection{Excess Timing Noise}

In Figure~\ref{fig5} the excess timing noise between MJD $51200$ and 
$53200$ coincides with the duration of maximum gradient of DM of
J1939. This manifests as a noisy bump when the fitted curve is subtracted
from the data in the two panels in Figure~\ref{fig5}. This is an achromatic 
excess timing noise, so it is independent of the DM. Further, the DM in
this duration is very well estimated due to excellent overlap of the 
digitized DM values and those estimated in this work. Currently the origin
of this excess timing noise is unknown.

\section{Discussion}

\cite{Shannon2013} studied the timing noise of J1939 using $26$ years of timing 
data. They have excluded the possibility of a single object causing these timing
results because of ``the lack of an obvious periodicity" in their timing residuals; 
see first line of first para after their Eq. $2$ in their section $2$. This
situation does not apply to this work, which has ~5 more years of data, and has a 
clear periodicity. Moreover, \cite{Shannon2013} have themselves flagged the most 
serious drawback of their asteroid belt model, viz., "that it is difficult, though 
not impossible, to test"; see first line of their section 7.

Before discussing the planetary companion/precession models, a few technical 
issues will be highlighted. 

In this work, only one cycle of either planetary orbit or precession is
available for analysis. This would limit the confidence with which the
corresponding parameters can be estimated. Unfortunately, even the second 
cycle of data is unlikely to be obtained any time soon.

Assuming a braking index of 3, the $\ddot \nu$ expected for J1939 is $8.8 
\times 10^{-30}$ Hz sec$^{-2}$. The total duration of observation of 
$11324$ days would cause $8.8 \times 10^{-30} \times (11324 * 86400)^3 / 
6.0 = 0.0014$ of additional phase across the entire duration, which is 
about $2.2$ $\mu$s. This is negligible compared to the $\approx 130$ 
$\mu$s amplitude of the sinusoidal timing noise. So the sinusoids in 
Figures \ref{fig5} and \ref{fig6} are not an artifact of the cubic term.

The fact that the period of the timing noise lies very close to the total 
duration of observation could cause some concern. However, there appear
to be no known artifacts in TEMPO2 (or in any other software/algorithm
that is used here) that might mimic a periodicity of the order of the
length of the data used. This issue is discussed in greater detail in
the Appendix.

Why does the MCMC algorithm produce much smaller errors than TEMPO2? I
believe it is because the MCMC planetary companion modeling involves a 
non-linear fit to an exact ellipse, while the TEMPO2 binary fit involves 
a linear fit to an approximate ellipse (an ellipse that has been 
linearized with respect to its parameters). Two binary models (BT and DD) 
were tried within TEMPO2, and both failed to converge most of the time. 
Convergence occurred when the initial values were almost the converged 
values, but the errors on the parameters were larger; here many more 
cycles of the sinusoid in the data would have helped.
This issue is discussed in greater detail in the Appendix.

Now,  the important question concerning the planetary companion of J1939 is --
how did it form around a millisecond pulsar (MSP)? \citet{Phillips1994} summarize the various 
possibilities. In a few of these, the planet is formed before the neutron star 
(NS) is formed, and somehow survives the supernova explosion (SNE); but in 
most of them the planet is formed after the NS.

In the former scenarios, a planet that is formed around a normal star, and that 
survives the passage through the expanding Red Giant envelope of the star, would still 
become unbound after the SNE since more than half the mass of the system might 
be lost to the ISM. This can be avoided only if the SNE is asymmetric and 
preferentially oriented with respect to the velocity of the planet, or 
alternately, if the planet's orbit is highly eccentric. In either case, this 
scenario may work for a slow pulsar but not for a MSP, which 
has to be spun up to ms periods by accretion from a companion star.  Another 
possibility is that the planet formed around a system of normal binary stars, 
one of which underwent a SNE, which did not disrupt the binary because less 
than half the mass of the system got expelled into the ISM, and later the NS 
spiraled into the companion star. Finally the simplest scenario, but 
statistically the least probable, would be the capture of a planet around a 
normal star by a MSP in a chance exchange interaction (see references in 
\citet{Phillips1994}). 

In the latter scenarios, the planet is formed from the disk material around
the MSP. But to spin a NS to ms periods one requires mass accretion 
from a companion star,
which must somehow be gotten rid of later, leaving just the disk material. One 
mechanism of doing this is through evaporation of the star by the pulsar 
wind; some of the evaporated material forms the disk from which the planet
can form. This mechanism is expected to form planets that are approximately Moon
size, so this scenario appears to be a possibility in J1939. Yet another 
scenario is that the companion of the NS is a white dwarf (WD), and the mass
losing WD is reduced to a disk (see references in \citet{Phillips1994}).

While it is not clear which of these various possibilities (and the several 
more summarized by \citet{Phillips1994}) explain the case of J1939, this work 
places the following constraints: (1) the planet around MSP J1939 is 
at a distance of $11$ AU, which is similar to the distance of 
$10.26$ AU of the planet around the slow pulsar B0329+54 
\citep{Starovoit2017}, and much larger than the distances of $0.36$ and 
$0.47$ AU of the two planets around the MSP B1257+12 
\citep{Wolszczan1992}; (2) its eccentricity $e = 0.21$ is similar to that of the
planet around B0329+54 ($0.24$), while the eccentricities of the two planets
of B1257+12 are almost negligible ($\approx 0.02$); and (3) the masses of 
the three planets mentioned above are $2$, $3.4$ and $2.8$ Earth masses, 
respectively, while the planet around J1939 is about $100$ times less 
massive. Thus, while J1939 shares an evolutionary scenario with the ms PSR 
B1257+12 in terms of mass accretion, its planetary distance and eccentricity 
appear to be similar to 
that of the slow PSR B0329+54, whose evolution is entirely different.
As an illustration of the constraints, theories of planet formation around 
B1257+12 have to invoke mechanisms to circularize the planets' orbits, either
during their formation or later, while
theories of planet formation around J1939 must suppress the very same 
mechanisms, while starting off with the common scenario of mass accretion 
that is mandatory for MSPs.

Now coming to the precession model of J1939, in this work precession under the 
influence of the electromagnetic torque of the pulsar is considered, not 
free precession. The main difference between the two cases is (1) the 
timing noise would be strictly a sine wave in the latter case, while it 
will have a second harmonic in the former case; and (2) the amplitude of
the sine wave will be significantly enhanced when torque drives the 
precession, even if the oblateness $\epsilon$ is very small 
\citep{Jones2001, Link2001, Akgun2006}.

Next, the shape of J1939 is assumed to be bi-axial for which two of the
moments of inertia are equal. However, given the very small value of
$\epsilon$ estimated in this work, one should also explore the tri-axial 
case ($I_3 > I_2 > I_1$).  \cite{Akgun2006} have derived formulae for 
the timing noise in this case, which are very complicated, and whose 
application is beyond the scope of this work.

The oblateness J1939 is $\epsilon \approx 1.57 \times 10^{-12}$; in 
comparison, the value for the Crab pulsar is (assuming that it 
precesses) $\epsilon \approx 6.27 \times 10^{-10}$, for PSR B1642-03 it
is $\epsilon \approx 4.48 \times 10^{-9}$ \cite{Jones2001}. The latter
two pulsars are not MSPs, so it is interesting to speculate if
the very low oblateness of J1939 has something to do with its being
recycled due to accretion. Could this process have kept the surface
of J1939 very hot for so long that the NS adjusted to a new equilibrium 
shape having very low oblateness? Here it should be noted that there are
two contributions to the oblateness -- a centrifugal deformation due to
the very high rotation of J1939, and Coulomb deformation due to the 
rigidity of its crust. Precession depends only upon the latter (see
section $3$ of \citet{Link2001}). Thus the very low $\epsilon$ of J1939
would imply that its Coulomb crust is not at all strained (see section 
$6$ of \citet{Link2001}).

Precession in J1939 can be damped if its crust couples to the interior
super fluid, on time scales of $2 \pi \tau_f \nu$ precession periods, 
where $\tau_f$ is the coupling time scale and $\nu$ is the rotation
frequency of J1939 (see section $3$ of \citet{Link2001}). Since 
it is not damped in J1939 for $31$ years, it implies that
the crust of J1939 is essentially decoupled from its super fluid 
interior. This would imply that either super fluid vortices do not
pin to the crust of J1939, or its precession is strong enough to
break the pinning. Either way one would not expect to see timing 
glitches in J1939, since that involves sudden unpinning of pinned
vortices \citep{Alpar1984}.

That J1939 has displayed no glitches so far is consistent with the 
belief that pinning of super fluid vortices suppresses precession 
in pulsars \citep{Shaham1977}.

Finally, the method of combining the IPTA and NANOGrav data adopted here might
prove useful to PTAs in extending their search to lower spatial frequencies.

The treatment of red noise in this paper can probably be done in an alternate way
using modern statistical techniques such as ARMA, ARIMA, ARFIMA, GARCH \citep{Feigelson2018}.

\acknowledgments

I thank M. T. Lam  and J. P. W. Verbiest for tips regarding analysis of NANOGrav 
and IPTA data, respectively, and discussion. I thank M. T. Lam for detailed
discussion regarding several technical aspects of this manuscript. I thank the 
Statistics Editor for bringing to my attention the work of \cite{Feigelson2018}.
I thank the referee for useful discussion and suggestions.

\software{TEMPO2 \citep{Hobbs2006, Edwards2006}, 
PyMultiNest \citep{Buchner2014}, 
Scipy \citep{JonesOliphant2001}, 
MULTINEST \citep{Feroz2008, Feroz2009, Buchner2014}
}

\section{Appendix}

\subsection{TEMPO2 Usage}

\begin{table}[h]
\scriptsize
\setlength\tabcolsep{1.5pt}
\hspace*{-1.0cm}
	\caption{Some TEMPO2 parameters used by IPTA, NANOGrav and this work.  \label{tabl3}}
\begin{tabular}{lccc}
\hline
\hline
{PARAMETER} & {IPTA}  & {NANOGrav} & {THIS WORK} \\
\hline
	NE\_SW                 &   $4.0$   &   $0.0\footnote{SOLARN0 is used to set the zero value}$   &   $4.0$ \\
EPHVER                  &   $5$   &   &   $5$ \\
EPHEM                   &   DE421   &   DE436   &   DE436 \\
CLK                     &   TT(BIPM2013)   &   TT(BIPM2015)   &   TT(BIPM2015) \\
UNITS                   &   TCB   &   TDB   &   TCB \\
TIMEEPH                 &   IF99   &   FB90   &   IF99 \\
T2CMETHOD               &   IAU2000B   &   TEMPO   &   IAU2000B \\
DILATEFREQ              &   Y   &   N   &   Y \\
PLANET\_SHAPIRO        &   Y   &   N   &   Y \\
CORRECT\_TROPOSPHERE   &   Y   &   N   &   Y \\
\end{tabular}
\end{table}

Table~\ref{tabl3} shows some important parameters of the TEMPO2 runtime environment
that differ for the IPTA and the NANOGrav groups. The NANOGrav group uses a more modern
planetary ephemeris (EPHEM) and a more modern realization of the Terrestrial Time 
(CLK). However, the NANOGrav group uses an older method of transforming the observatory 
coordinates to the celestial frame for ``Roemer" delay (T2CMETHOD), an older method 
of conversion from SAT to BAT (UNITS), and also an older method of estimating the 
``Einstein" delay (TIMEEPH). They also do not apply gravitational red shift and time 
dilation to observing frequency (DILATEFREQ), do not apply tropospheric delay 
corrections (CORRECT\_TROPOSPHERE), and do not compute Shapiro delay due to the 
planets in the solar system (PLANET\_SHAPIRO). Finally, they do not compute the
dispersion delay in the solar system due to the solar wind; they set the solar
electron density (at $1$ AU) to zero (NE\_SW).

\subsection{Timing Noise}

\begin{figure}[h]
\centering
\hspace*{-1.0cm}
\includegraphics[width=1.2\hsize]{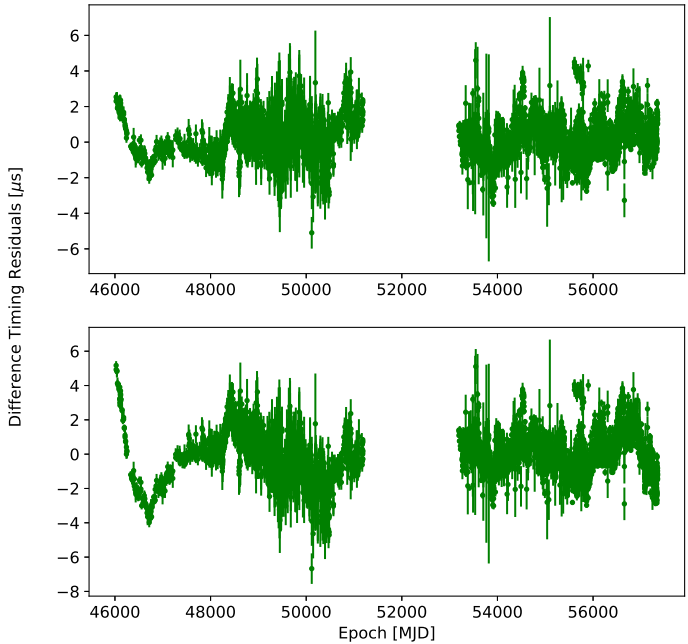}
\caption{Difference between the data and the modeled curves
         of Figure~\ref{fig5}.
	}
\label{fig8}
\end{figure}

Figure~\ref{fig5} shows the timing residuals fit to a planetary companion model
using the IPTA (top panel) and the NANOGrav (bottom panel) methods of correction 
for “red” noise correlation. The green curves are the data while the red dashed
curves represent the planetary model. Figure~\ref{fig8} shows the corresponding plots after 
subtracting the model from the data; the excess timing noise between MJD 51200 
and 53200 has been ignored. The standard deviations of the difference timing 
residuals in the top and bottom panels of Figure~\ref{fig8} are $1.0$ $\mu$s 
and $1.2$ $\mu$s, respectively, both of which are comparable to the value $0.5$ 
$\mu$s  estimated in Figure~\ref{fig1}. In both panels of Figure~\ref{fig8} the
fit appears to be poor only in the initial $1000$ days of the data; for the 
rest of the data the differences are flat within errors. Therefore one concludes 
that the planetary companion model is a genuine representation of the timing
residuals of J1939.

\begin{figure}[h]
\centering
\hspace*{-1.0cm}
\includegraphics[width=1.2\hsize]{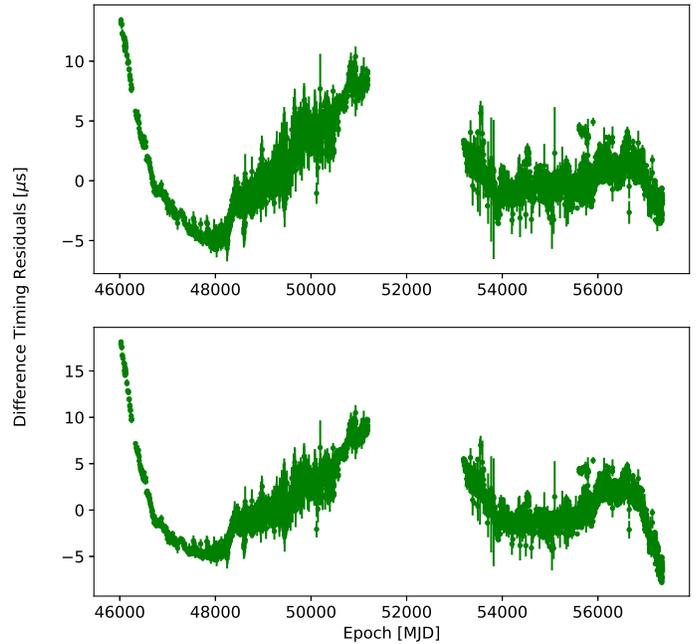}
\caption{Difference between the data and the modeled curves
         of Figure~\ref{fig6}.
	}
\label{fig9}
\end{figure}

Figure~\ref{fig9} shows difference timing residuals of the data in Figure~\ref{fig6},
in which a precession model has been fit. The standard deviations of the difference 
timing residuals in the top and bottom panels of this figure are $2.5$ $\mu$s and
$2.9$ $\mu$s, respectively. These are significantly higher than those in 
Figure~\ref{fig8}; the fit is poor over the first half of the data. This supports 
the contention in the paper that the planetary model is a better fit to the timing 
residuals of J1939 than the precession model. However the latter can not be ruled out 
entirely since it does fit the later half of the data in the top panel of 
Figure~\ref{fig6}, the standard deviation of the fit for this data being $1.2$ $\mu$s.

An important issue concerns the spread in period ($P$) values in Tables $1$ and $2$.
The mean value of the periods of the four planetary orbits in Table~\ref{tabl1} is
$11739$ days, while their standard deviation is $2.1$ years, which is significantly 
larger than the formal errors on the periods, which are of the order of a few days. 
These are clearly due to uncertainties in the red noise models as well as the choice 
of the T2 parameters. From a practical point of view the larger uncertainty should 
be used in understanding the period of the planetary orbits. Similarly, the mean and
standard deviation of the periods of the four precession models in Table~\ref{tabl2}
are $11989$ days and $2.4$ years, respectively.

As mentioned in the text, the large values of reduced $\chi^2$ in Tables $1$ and $2$
reflect the fact that the use of the T2 parameters is of limited utility for the
data of J1939. In any case such large values of reduced $\chi^2$ should not be a
surprise when combining highly disparate data from $36$ different sub-systems.

\begin{figure}[h]
\centering
\hspace*{-1.0cm}
\includegraphics[width=1.2\hsize]{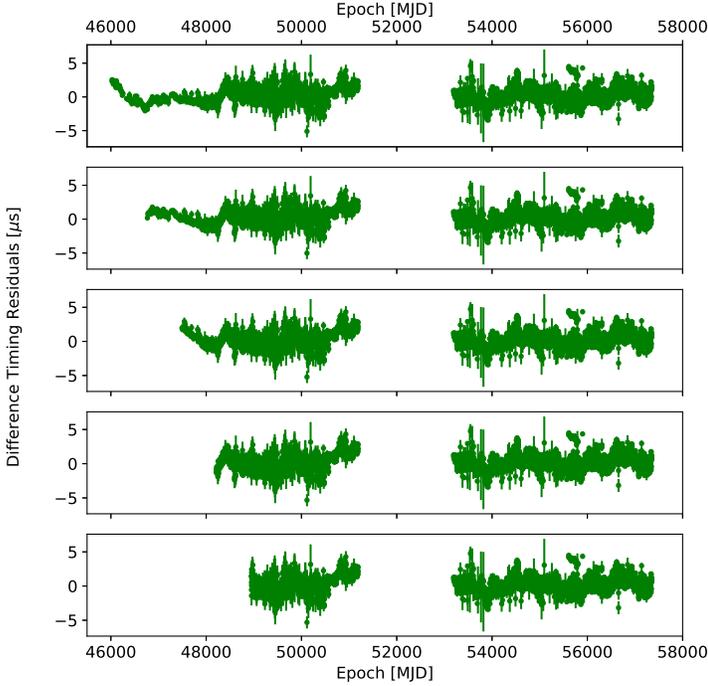}
\caption{Difference between the data and the modeled curves
         of top panel of Figure~\ref{fig5}. Top panel: Same as
	 the top panel of Figure~\ref{fig5}. Subsequent panels: 
	 Each panel has two years of data less than the previous 
	 panel.
	}
\label{fig10}
\end{figure}

Finally, it is relevant to ask if there are correlations between the length of
data used and the periodicities derived, in Figure~\ref{fig5}. This might be 
the case if the timing residuals were due to random variations in the frequency 
$\nu$, the variations having a steep power law spectrum. The top panel of 
Figure~\ref{fig10} is exactly the same as the top panel of Figure~\ref{fig8}; 
this is for better comparison with the rest of the panels of Figure~\ref{fig10}.
The second panel (from the top) in Figure~\ref{fig10} is the same, but with the 
first two years of data excised; the next panel has the first four years of data 
excised; the last two panels have the first six and eight years of data excised, 
respectively. The planetary model fits the smaller amount of data equally well. 
The corresponding derived parameters are given in Table~\ref{tabl4}.

\begin{table}[t]
\scriptsize
\setlength\tabcolsep{1.5pt}
\hspace*{-1.0cm}
\caption{Results of fitting the planetary companion model to the timing noise, 
	using the IPTA methods of correction for ``red" noise correlation, and
	using local T2 parameters. The second column of this table is identical
	to the third column of Table~\ref{tabl1}. Columns three on wards display
	the parameters obtained when decreasing amount of data is used for the 
	fit; the number of years of data excised from the starting epoch is 
	displayed at the head of each column.
	\label{tabl4}}
\begin{tabular}{lccccc}
\hline
\hline
\multicolumn{2}{c}{Method = IPTA} & \multicolumn{2}{c}{T2param = local} \\
\hline
               & $0$ yr            & $2$ yr            & $4$ yr            & $6$ yr            & $8$ yr \\
\hline
{$A$} ($\mu$s) & $197.6 \pm 0.1$   & $197.8 \pm 0.1$   & $198.8 \pm 0.1$   & $199.3 \pm 0.2$   & $200.0 \pm 0.7$ \\
{$e$}          & $0.158 \pm 0.001$ & $0.152 \pm 0.001$ & $0.147 \pm 0.001$ & $0.145 \pm 0.001$ & $0.145 \pm 0.01$ \\
{$P$} (days)   & $13068 \pm 3$     & $13139 \pm 6$     & $13276 \pm 13$    & $13337 \pm 21$    & $13370 \pm 51$ \\
{rms} ($\mu$s) & $1.0$             & $1.0$             &  $1.0$            & $1.0$             & $1.0$ \\
{$\chi^2_d$}   & $30.2$            & $29.5$            &  $29.1$           & $29.8$            & $32.5$ \\
\end{tabular}
\end{table}

Columns two to six of Table~\ref{tabl4} correspond to the best fit parameters of panels one
to five of Figure~\ref{fig10}, respectively. Column two of this table is identical to column
three of Table~\ref{tabl1}. Columns three to six of Table~\ref{tabl4} have two, four, six 
and eight years of initial data excised before fitting the planetary companion model. It is
clear that the excised data produces results that are consistent with the complete data. The 
rms in columns three to six are almost the same (up to the first decimal place) as the rms in
column 2. So are the $\chi^2$ per degree of freedom. The parameters $A$, $e$ and $P$ in 
columns three to six of Table~\ref{tabl4} are also almost similar to those in column two, 
with a tendency to have larger formal errors when more data is excised. Therefore there 
appears to be no correlation of the derived periods $P$ with the length of the data used for 
the planetary model. This is reflected in the almost similar difference timing residuals in 
the panels of Figure~\ref{fig10} for the data that is common across panels. 


\vfill \eject

\end{document}